\documentclass[12pt]{article}
\usepackage{graphicx}
\usepackage{color}
\usepackage{cite}
\def\downstrut{\vrule height 1ex depth 1.0ex width 0pt}
\def\upstrut{\vrule height 2.5ex depth 0.0ex width 0pt}

\def \beq{\begin{equation}}
\def \eeq{\end{equation}}
\def\eqref#1{(\ref{#1})}
\def\bea{\begin{eqnarray}}
\def\eea{\end{eqnarray}}
\def\jpsi{\hbox{$J\kern-0.2em/\kern-0.1em\psi$}}

\def \s{\sqrt{2}}

\def\URLtilde{\lower0.2em\hbox{$\tilde{\phantom{a}}$}}
\def\mycomm#1{\hfill\break\strut\kern-3em{\color{red}\tt ====> #1
\color{black}}\hfill\break}
\newcount\timecount
\newcount\hours \newcount\minutes  \newcount\temp \newcount\pmhours
\hours = \time
\divide\hours by 60
\temp = \hours
\multiply\temp by 60
\minutes = \time
\advance\minutes by -\temp
\def\hour{\the\hours}
\def\minute{\ifnum\minutes<10 0\the\minutes
\else\the\minutes\fi}
\def\clock{
\ifnum\hours=0 12:\minute\ AM
\else\ifnum\hours<12 \hour:\minute\ AM
\else\ifnum\hours=12 12:\minute\ PM
\else\ifnum\hours>12
\pmhours=\hours
\advance\pmhours by -12
\the\pmhours:\minute\ PM
\fi
\fi
\fi
\fi
}

\def\monthname{\relax\ifcase\month 0/\or January\or February\or
March\or April\or May\or June\or July\or August\or September\or
October\or November\or December\else\number\month/\fi}

\def\bold#1{\setbox0=\hbox{$#1$}     \kern-.025em\copy0\kern-\wd0
\kern.05em\copy0\kern-\wd0
\kern-.025em\raise.0433em\box0 }

\begin{document}
\setcounter{footnote}{1}
\rightline{EFI 16-1}
\rightline{TAUP 3005/16}
\rightline{arXiv:1601.nnnnn}
\vskip1cm

\begin{center}
{\Large \bf Exotic resonances due to $\eta$ exchange}
\end{center}
\bigskip

\centerline{\bf Marek Karliner$^a$\footnote{{\tt marek@proton.tau.ac.il}}
 and Jonathan L. Rosner$^b$\footnote{{\tt rosner@hep.uchicago.edu}}}
\medskip

\centerline{$^a$ {\it School of Physics and Astronomy}}
\centerline{\it Raymond and Beverly Sackler Faculty of Exact Sciences}
\centerline{\it Tel Aviv University, Tel Aviv 69978, Israel}
\medskip

\centerline{$^b$ {\it Enrico Fermi Institute and Department of Physics}}
\centerline{\it University of Chicago, 5620 S. Ellis Avenue, Chicago, IL
60637, USA}
\bigskip
\strut

\begin{center}
ABSTRACT
\end{center}
\begin{quote}
The meson $X(3872)$ and several related states appear to be, at least in part,
hadronic molecules in which a heavy flavored meson (such as $D^0$) is bound
to another heavy meson (such as $\bar D^{*0}$).  Although not the only effect
contributing to the binding, pion exchange seems to play a crucial role in
generating the longest-range force between constituents.  Mesons without $u$
and $d$ light quarks (such as $D_s$) cannot exchange pions, but under suitable
circumstances can bind as a result of $\eta$ exchange.  Channels in which this
mechanism is possible are identified, and suggestions are made for searches
for the corresponding molecular states, including a manifestly exotic baryonic
$\Lambda_c \bar D_s^*$ resonance decaying into $\jpsi\,\Lambda$.
\end{quote}

\smallskip

\leftline{PACS codes: 12.39.Hg, 12.39.Jh, 14.20.Pt, 14.40.Rt}
\bigskip


The discovery more than a dozen years ago of an extremely narrow resonance,
$X(3872)$ \cite{X3872}, right at the $D \bar D^*$ threshold, inaugurated a
flurry of observations of charmonium-like and bottomonium-like resonances
similarly correlated with thresholds.  A number of these could be identified
as possessing a significant ``molecular'' component, in which a heavy
charmed or bottom hadron was bound to an anticharmed or anti-bottom hadron
\cite{Voloshin:1976ap,DeRujula:1976qd}.  When these hadrons possess light
quarks, the longest-range force between them is single-pion exchange, in
analogy with the deuteron which binds via exchange of pions and other light
mesons \cite{deusons,Tornqvist:2004qy,Thomas:2008ja,Suzuki:2005ha,Fleming:%
2007rp,Karliner:2015ina}.  The question then arises as to whether a related
mechanism can play a role in binding heavy hadrons which contain no $u,d$
quarks.  In this note we identify potential channels in which $\eta$
exchange is the longest-range force, and can thus form bound states with
quark content such as $(c \bar s)(\bar c s)$.  We predict masses based on
the proximity to thresholds of charmed-antistrange and anticharmed-strange
pairs.  Such a proximity is a widespread feature of S-wave structures
\cite{Rosner:2006vc}.
\medskip

There have been observations \cite{Aaltonen:2009tz,Aaltonen:2011at,%
Chatrchyan:2013dma,Abazov:2013xda,Abazov:2015sxa} or failures to observe
\cite{Shen:2009vs,Aaij:2012pz,Lees:2014lra} a $\jpsi\, \phi$ resonance at 4140
MeV, which does not correspond to any known $D_s^{*+} D_s^{*-}$ threshold.
Both $\eta$ and $\phi$ exchange were considered in a work identifying the 4140
MeV state as a $D_s^{*+} D_s^{*-}$ molecule \cite{Liu:2009ei}, with predicted
$J^P = 0^+$ and $2^+$ masses highly dependent on an arbitrary cutoff parameter.
Such a molecule was also considered in Ref.\ \cite{Ding:2009vd}, where the
binding was due to $\eta$, $\sigma$, and $\phi$ exchange.  The large binding
energy in these two works is somewhat suspicious in view of the short range
of these potentials.  A recent work explains the 4140 MeV state as a mixture of
10\% $D^{*0} \bar D^{*0}$, 10\% $D^{*+} D^{*-}$, and 80\% $D_s^{*+} D_s^{*-}$
\cite{Chen:2015fdn}.
If the existence of the $\jpsi\,\phi$ resonance at 4140 MeV is confirmed,
it is likely to be due to an additional mechanism, beyond the $\eta$ exchange 
discussed here.
\medskip

\begin{table}
\caption{Possible S-wave resonances with two $D_s$ mesons below 5 GeV.
\label{tab:ds}}
\begin{center}
\begin{tabular}{c c c c c} \hline \hline
\upstrut
    States ($J^P$) & $M$ & $M-M(\jpsi)$ & Binding   & Allowed \\
\downstrut       & (MeV) &  $-M(\phi)$  & by $\eta$? & $J^P$ \\
\hline
\upstrut
      $D^+_s(0^-)~D^-_s(0^-)$    &  3936.6  & --179.8 &  No &   --  \\
    $D^+_s(0^-)~D^{*-}_s(1^-)$   &  4080.4  & --36.0  & Yes & $1^+$ \\
  $D^{*+}_s(1^-)~D^{*-}_s(1^-)$  & 4224.2 &  107.8  & Yes & $0^+,2^+{}^{~a}$ \\
  $D^+_s(0^-)~D^{*-}_{s0}(2317)(0^+)$  &  4286.0  &  169.6  & Yes & $0^-$ \\
$D^+_s(0^-)~D^-_{s1}(2460)(1^+)$ & 4427.8  & 311.4  &  No$^b$ & $[1^-]^{~b}$ \\
$D^{*+}_s(1^-)~D^{*-}_{s0}(2317)(0^+)$&4429.8& 313.4 & No$^b$ & $[1^-]^{~b}$ \\
 $D^+_s(0^-)~D^-_{s1}(2536)(1^+)$&  4503.4  &  387.0  &  No  &  --  \\
 $D^+_s(0^-)~D^{*-}_{s2}(2573)(2^+)$   &  4540.2  &  423.8  & Yes & $2^-$ \\ 
$D^{*+}_s(1^-)~D^-_{s1}(2460)(1^+)$& 4571.6 &  455.2  & Yes & $0^-,1^-,2^-$ \\
$D^{*+}_{s0}(2317)(0^+)~D^{*-}_{s0}(2317)(0^+)$& 4635.4 &  519.0  &  No & -- \\
$D^{*+}_s(1^-)~D^-_{s1}(2536)(1^+)$& 4647.2 &  530.8  & Yes & $0^-,1^-,2^-$ \\
$D^{*+}_s(1^-)~D^{*-}_{s2}(2573)(2^+)$ & 4684.0 & 567.6 & Yes & $1^-,2^-,3^-$\\
$D^{*+}_{s0}(2317)(0^+)~D^-_{s1}(2460)(1^+)$&4777.2& 660.8  & Yes & $1^+$ \\
$D^{*+}_{s0}(2317)(0^+)~D^-_{s1}(2536)(1^+)$&4852.8$^c$&736.4& Yes & $1^+$ \\
$D^{*+}_{s0}(2317)(0^+)~D^{*-}_{s2}(2573)(2^+)$&4889.6$^c$& 773.2 &  No &  --    \\
$D^+_{s1}(2460)(1^+)D^-_{s1}(2460)(1^+)$ & 4919.0$^c$ & 802.6 & Yes &
 $0^+,2^+{}^{~a}$\\
$D^+_{s1}(2460)(1^+)D^-_{s1}(2536)(1^+)$ & 4994.6$^c$ & 878.2 & Yes &
 $0^+,1^+,2^+$ \\
\hline \hline
\end{tabular}
\end{center}
\leftline{$^a$ $J^P = 1^+$ forbidden by symmetry.}
\leftline{$^b$ Proximity of these two channels may lead to binding.  See text.}
\leftline{$^c$ Cannot be produced in $B \to K X$ because of kinematic mass
limit.}
\end{table}

\begin{figure}
\begin{center}
\includegraphics[width = 0.98\textwidth]{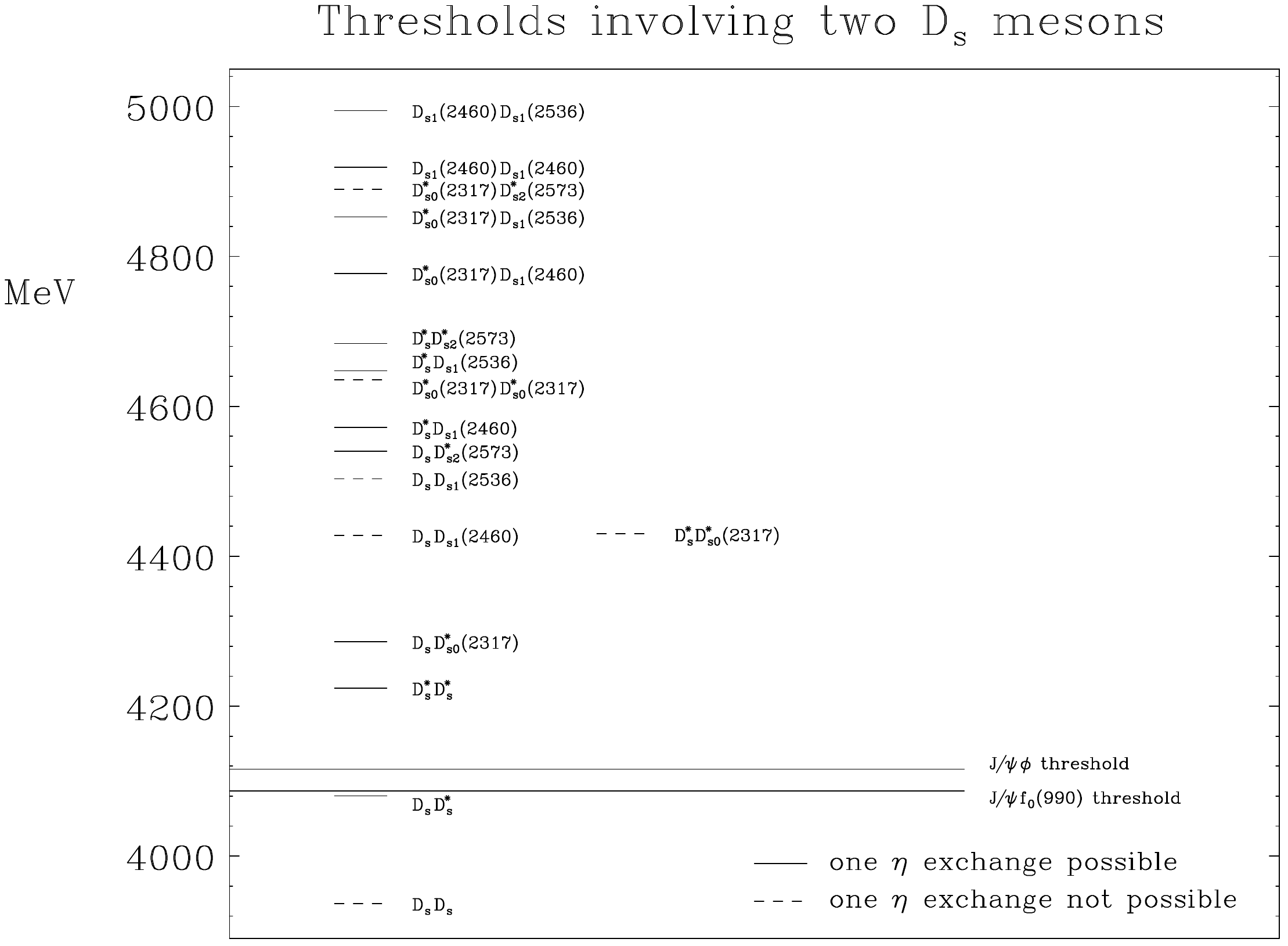}
\end{center}
\caption{Comparison of $D_s^{(*)+} D^{(*)-}_s$ thresholds with those of
$\jpsi\, f_0$ and $\jpsi\, \phi$.
\label{fig:ds}}
\end{figure}

The pseudoscalar $\eta$ cannot couple to a pair of scalar or pseudoscalar
mesons.  Thus some $(c \bar s) (\bar c s)$ channels will receive a contribution
to their binding from $\eta$ exchange, while others will not.  In Table
\ref{tab:ds} we summarize possible resonances involving two $D_s$ mesons,
with special attention to those which can be produced in decays of the form
$B \to K X$, i.e., states below about 4786 MeV.  We take the masses
$M(D_s) = 1968.3$ MeV, $M(D^*_s) = 2112.1$ MeV, $M(D^*_{s0}(2317)) = 2317.7$
MeV, $M(D_{s1}(2460)) = 2459.5$ MeV, $M(D_{s1}(2536)) = 2535.11$ MeV,
$M(D^*_{s2}(2573)) = 2571.9$ MeV, $M(\jpsi) = 3096.92$ MeV, $M(\phi)
= 1019.46$ MeV, and $M(f_0) = 990$ MeV from Ref.\ \cite{PDG}.  Thresholds
involving two $D_s$ mesons are compared with the $\jpsi \,f_0$ and $\jpsi\,
\phi$ thresholds in Fig.\ \ref{fig:ds}.
\bigskip

We now discuss the sign of the forces due to $\eta$ exchange in some of the 
lowest-mass channels in which binding is possible.
\medskip

(i) $D^+_s~D^{*-}_s$:  This channel is analogous to $D^0~\bar D^{*0}$ if one
replaces a $u$ or $\bar u$ quark with an $s$ or $\bar s$ quark.  Hence the
binding due to $\eta$ exchange for the $C=+$ combination $(D^+_s~D^{*-}_s +
D^{*+}_s~D^-_s)/\s$ should be of the same sign as it is for the $X(3872)$,
which is generally acknowledged as having a significant component of the
$C=+$ combination $(D^0~\bar D^{*0} + D^{*0}~\bar D^0)/\s$.  The range, of
course, will be smaller by a factor of $m_\pi/m_\eta$ than it is for pion
exchange.  As the $D^+_s~D^{*-}_s$ threshold is 36 MeV below $M(\jpsi) +
M(\phi)$, and just below $M(\jpsi) + M(f_0)$, the most one can expect is
an enhancement in the $M_{\jpsi \phi}$ and $M_{\jpsi f_0}$ spectra near
threshold.
\medskip

(ii) $D^{*+}_s~D^{*-}_s$:  The related channel $D^* \bar D^*$ was analyzed
in Ref.\ \cite{Karliner:2015ina}, where it was concluded that the most
attractive channel was the one with $I=J=0$.  This was a consequence of the
expectation values
\beq
\langle I_1 \cdot I_2 \rangle = [1/2][I(I+1) - I_1(I_1+1) - I_2(I_2+1)]
= (-3/4,+1/4)~{\rm for}~I=(0,1)~,
\eeq
\beq
\langle J_1 \cdot J_2 \rangle = [1/2][J(J+1) - J_1(J_1+1) - J_2(J_2+1)]
= (-2,-1,+1)~{\rm for}~J=(0,1,2)~,
\eeq
where the most attractive channel for a $q \bar q$ interaction is the one
with the {\it largest} value of $\langle I_1 \cdot I_2~J_1 \cdot J_2 \rangle$
\cite{deusons}.  In the present case, in which the isospin factor is absent,
the most attractive channel will be that with $J=2$.  Thus, $\eta$
exchange between $D^{*+}_s$ and $D^{*-}_s$ should give rise to a $J^P = 2^+$
resonance near 4224 MeV decaying to $\jpsi~\phi$.
\medskip

(iii) $D^+_s~D^{*-}_{s0}(2317)$:  The forces due to $\eta$ exchange will be
equal and opposite for eigenstates of the matrix
\beq \label{eqn:2x2}
V \sim \left[ \begin{array}{c c} 0 & -1 \cr -1 & 0 \end{array} \right]
\eeq
in the channels $[D^+_s~D^{*-}_{s0}(2317),D^+_{s0}(2317)~D^{*-}_s]$ (cf.\ the discussion
of $D \bar D^*$ in Ref.\ \cite{Karliner:2015ina}).  The eigenstates have
positive and negative $C$, and thus $J^{PC} = 0^{- \pm}$.  The attractive
channel, with $C=+$, can decay to $\jpsi~\phi$.  One would then see a resonance
near 4286 MeV with $J^{PC} = 0^{-+}$ decaying to $\jpsi~\phi$.
Indeed, the CDF Collaboration has $3.1\sigma$ evidence for a state at
$4274.4^{+8.4}_{-6.7} \pm 1.9$ MeV decaying to $\jpsi~\phi$
\cite{Aaltonen:2011at}, identified as a $D^+_s~D^{*-}_{s0}(2317)$ molecule in
Refs.\ \cite{He:2011ed} and \cite{Finazzo:2011he}.
\medskip

(iv) $D^+_s~D^-_{s1}(2460)$ and $D^{*+}~D^{*-}_{s0}(2317)$:  The proximity of
these two channels means that mixing between them due to $\eta$ exchange may be
possible, with an interaction of the form (\ref{eqn:2x2}).  One should then
expect a $J^P = 1^-$ resonance near 4429 MeV decaying to $\jpsi~\phi$.  The
mixing will produce two eigenstates of opposite $C$, with $V$ attractive in the
$C=+$ channel.
\medskip

(v) We have included $D^*_{s2}(2573)$ in the discussion even though it is not as
narrow as the other states, having a width of $17 \pm 4$ MeV.  Any resonance
involving it will be at least as broad, such as the predicted state around
4540 MeV with $J^P = 2^-$.  The potential is again of the form (\ref{eqn:2x2}),
with the lower-lying eigenstate having $C=+$.
\medskip

(vi) Arguments similar to those in (iii) may be applied to states near 4572,
4647, 4684, and 4777 MeV.  In each case $\eta$ exchange gives an attractive
force in one or more channels with $C=+$, giving resonances which can decay to
$\jpsi~\phi$.
\medskip

If it turns out that $\eta$ exchange can indeed lead to $D_s \bar D_s^*$
resonances, then analogous meson-baryon resonances should also exist,
by the same reasoning as in \cite{Karliner:2015ina}.
A prerequisite is that both the meson and the baryon must be heavy, and at
least one of them should not couple to pions.
The simplest example is a $\Lambda_c \bar D_s^*$ resonance, with quark 
content $c\bar c s u d$.  The relevant threshold is at 4398.6 MeV.
\medskip

If such a $\Lambda_c \bar D_s^*$ resonance does exist, its 
best chance of being formed is in
$\Lambda_b$ decay.  The decay $\Lambda_b \to \Lambda_c \bar D_s^*$
is Cabibbo favored. The mass of $\Lambda_b$ is 5619.5 MeV, so approximately 
1221 MeV needs to be carried off, e.g., by an extra $\pi^+ \pi^-$ pair
or, as recently suggested \cite{Feijoo:2015kts}, by an $\eta$.  
The $\Lambda_c \bar D_s^*$ resonance can decay through quark rearrangement
to $\jpsi\,\Lambda$, with $Q$-value of approximately 186 MeV.
The most promising discovery channel is then 
\beq
\Lambda_b \to \jpsi\, \Lambda\, (\pi^+ \pi^-~{\rm or}~\eta)
\label{Lambda_b-decay}
\eeq
where one looks for a $\jpsi\, \Lambda$ resonance around 4400 MeV.
\bigskip

When $u,d$ quarks are absent, $\eta$ exchange indeed seems to be the
longest-range single-particle-exchange force available to form hadronic
molecules of two systems containing heavy quarks.  It will be interesting
to see if the dynamics of this formation is sufficiently sensitive to $\eta$
exchange that the predicted states are observed.
\bigskip

We thank Tomasz Skwarnicki for many helpful comments on the manuscript.
The work of J.L.R. was supported in part by the U.S. Department of Energy,
Division of High Energy Physics, Grant No.\ DE-FG02-13ER41958.

\end{document}